\documentclass[aps,showpacs,twocolumn,prl]{revtex4}
\usepackage{epsfig}

\begin{document}

\title{Fracturing the optimal paths}

\author{J.~S.~Andrade Jr.$^{1,2}$, E.~A.~Oliveira$^{1}$, A.~A. Moreira$^{1}$ 
and H.~J.~Herrmann$^{1,2}$}

\affiliation{$^{1}$Departamento de F\'{\i}sica, Universidade Federal 
do Cear\'a, 60451-970 Fortaleza, Cear\'a, Brazil\\ $^{2}$IfB, HIF E12, 
ETH H\"onggerberg, 8093 Z\"urich, Switzerland}

\begin{abstract} 
Optimal paths play a fundamental role in numerous physical
applications ranging from random polymers to brittle fracture, from
the flow through porous media to information propagation. Here for the
first time we explore the path that is activated once this optimal
path fails and what happens when this new path also fails and so on,
until the system is completely disconnected.  In fact numerous
applications can be found for this novel fracture problem. In the
limit of strong disorder, our results show that all the cracks are
located on a single self-similar connected line of fractal dimension
$D_{b} \approx 1.22$. For weak disorder, the number of cracks spreads
all over the entire network before global connectivity is
lost. Strikingly, the disconnecting path (backbone) is, however,
completely independent on the disorder.
\end{abstract} 


\maketitle 

The identification and characterization of the optimal path in a
disordered landscape represents an important problem in theoretical
and computational physics as it can be intimately associated with many
relevant scientific and technological applications
\cite{Mezard84,Ansari85,Huse85,Kirkpatrick85,Kardar86,Kertesz92,Perlsman92,Havlin05} 
including brittle fracture, random polymers and transport in porous media. 
It has been shown that optimal paths extracted from energy landscapes
generated with weak disorder are self-affine and belong to the same
universality class of directed polymers \cite{Schwartz98}. In the strong
disorder or ultrametric limit, on the other hand, numerical works
\cite{Cieplak94,Porto97,Porto99} demonstrated the self-similar 
nature of the optimal path on two and three-dimensional lattices with
fractal dimensions given by $D_{f} \approx 1.22$ and $1.43$,
respectively.  
\begin{figure}[!htb]
\includegraphics*[width=8cm]{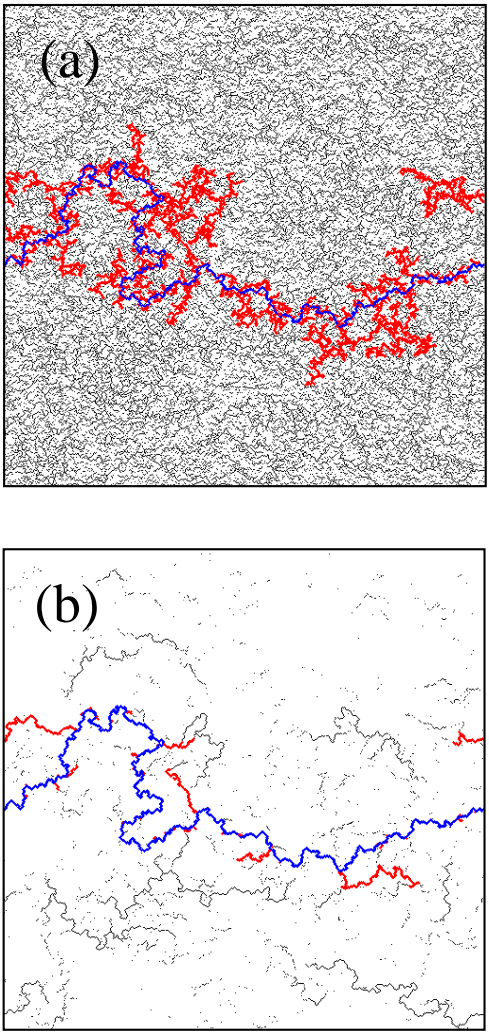}
\caption{(Color online) Typical realizations of the OPC model on a 
$512 \times 512$ lattice. In (a) the blocked sites generated under
weak disorder conditions ($\beta=0.002$) can be categorized in three
distinct types, namely, the loopless backbone of the OPC (shown in
blue) at which multiple dangling ends (shown in red) are attached, and
isolated clusters (shown in black) spread all over the network.
As shown in (b), under moderate disorder conditions ($\beta=6.0$), 
while the OPC backbone is preserved, the number of dangling ends
and isolated clusters is substantially reduced. Under conditions
of very strong disorder (not shown), only the OPC backbone remains.}
\end{figure}

Optimal paths can be chosen by nature as those of minimum energy as
occurs for instance in electrical or fluid flow through random media
or they can be chosen deliberately in man made devices to reduce a
cost function as for instance in internet or car traffic. Since these
paths are heavily used they are more likely prone to failure either by
overload, overheating or congestion. Two questions that naturally
arise are ({\it i}) how and when the system will eventually collapse
and ({\it ii}) how the topology and inhomogeneity of this fracture
affects its performance. It is the aim of the present Letter to
provide a novel modelization for this complex problem that captures
its essential features and gives some insight about the statistical
physics of these important questions. We first introduce a new model
to generate a macroscopic fault line, namely the Optimal Path Crack
(OPC) method, that is based on an iterative application of the
Dijkstra algorithm \cite{Dijkstra59} to two-dimensional random
landscapes. We then quantify the effect of disorder on the resulting
crack topology to substantiate the relevance of the OPC method as a
tool to understand the fracture of the random medium.

Our substrate is a square lattice of size $L$ with fixed boundary
conditions at the top and bottom, and periodic boundary conditions in
the transversal direction. Disorder is introduced in our model by
assigning to each site $i$ an energy value
$\epsilon_{i}=\exp[\beta(p_{i}-1)]$ \cite{Braunstein02}, where $p_{i}$
is a random variable uniformly distributed in the interval $[0,1]$,
and $\beta$ is a positive parameter that has the physical meaning of
inverse temperature. It can be readily shown that this transformation
is equivalent to choosing the values of $\epsilon$ from a power-law
distribution, namely $P(\epsilon_i)\sim{1/\epsilon_i}$, subjected to a
maximum cutoff given by $\epsilon_{max}=e^{\beta}$. The existence of
such a cutoff turns the hyperbolic distribution normalizable for any
finite value of $\beta$. In this way, $\beta$ represents the strength
of disorder since it controls the broadness of the energy
distribution.

The energy of any path in the system is defined here as the sum of all
the energies of its sites. In particular, the optimal path is the one
among all paths connecting the bottom to the top of the lattice that
has the smallest sum over all energies of its sites. This definition
is similar, but different from the definition of the shortest path in
a network \cite{Stauffer94}. Without loss of generality, since we only
consider positive $p_{i}$ values, the Dijkstra algorithm becomes a
suitable tool for finding the optimal path \cite{Dijkstra59}.

The OPC is formed as follows. Once the first optimal path connecting
the bottom and the top of the lattice is determined, we search for its
site having the highest energy which then becomes the first blocked
site, i.e., that can no longer be part of any path. This is equivalent
to impose an infinite energy to this site. Next, the optimal path is
calculated among the remaining accessible sites of the lattice, from
which the highest energy site is again removed, and so on. The blocked
sites can be viewed as ``micro-cracks'', and the process continues
iteratively until the system is disrupted and we can no longer find
any path connecting bottom to top.

In Fig.~1a we show the resulting spatial distribution of blocked sites
that constitutes the OPC in a typical random landscape generated under
weak disorder conditions ($\beta=0.002$). As depicted, the OPC
structure has three basic elements. Besides the loopless backbone of
the fracture (shown in blue) which effectively ``breaks'' the system
in two, we can observe the presence of dangling ends (shown in red) as
well as isolated clusters homogeneously distributed over the entire
network (shown in black). The situation becomes very different when we
increase the value of the disorder parameter $\beta$. As shown in
Fig.~1b, the amount of dangling ends and isolated clusters in a OPC
generated under moderate disorder conditions ($\beta=6.0$) becomes
significantly smaller than in the case of weak disorder. By increasing
further the value of $\beta$, finally only the backbone
remains. Interestingly this backbone is identical for all values of
$\beta$, while the entire set of blocked sites is highly dependent on
the way disorder is introduced in the system.
\begin{figure}[!htb]
\includegraphics*[width=8cm]{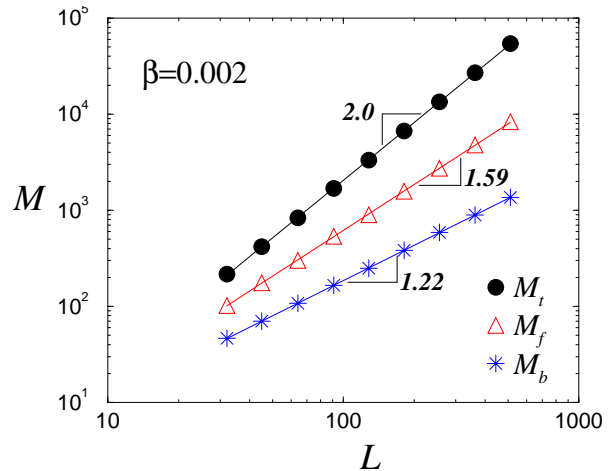}
\caption{(Color online) Logarithmic dependence on size of the mass of 
all blocked sites $M_{t}$ (circles) forming the OPC, the cluster of
all sites in the fracture $M_{f}$ (triangles), and the backbone mass
$M_{b}$ of the OPC that divides the system in two (stars), for
$\beta=0.002$. The system is considered to be in the weak disorder
regime for this value of $\beta$ and this range of system sizes. The
three solid lines are the least-squares fits to the data of
power-laws, $M_{t} \simeq L^{D_{t}}$, $M_{f} \simeq L^{D_{f}}$ and
$M_{b} \simeq L^{D_{b}}$, with exponents $D_{t}=2.00 \pm 0.01$,
$D_{f}=1.59 \pm 0.02$, and $D_{b}=1.22 \pm 0.02$, respectively.}
\end{figure}

The behavior shown in Fig.~1 is somehow related to the problem of
minimum path in disordered landscapes \cite{Porto99}. In that case,
the passage from weak to strong disorder in the energy distribution
reveals a sharp crossover between self-affine and self-similar
behaviors of the optimal path. In the strong disorder regime the
energy of the minimum path is controlled by the energy of a single
site. This situation occurs when the distribution of energies can not
be normalized, for instance, in the case of a power-law distribution,
$P(\epsilon_{i})\sim{\epsilon_{i}^{-\alpha}}$, for $\alpha\le{1}$. The
parameter $\beta$ alone, however, does not determine the limit between
weak and strong disorder, since this property also depends
significantly on the system size. More precisely, if $\beta$ is
sufficiently high, or the lattice size is sufficiently small, the
sampling of the distribution near the cutoff region is not so
relevant. For any practical purpose, this network is considered to be
in the strong disorder regime, resulting in a self-similar type of
scaling for the minimum path. By increasing the network size, we may
reach the point where one should expect to start sampling larger
energy values that are beyond the cutoff of the distribution. Above
this scale, the system will return to the weak disorder regime,
leading to a self-affine behavior for the minimum path. In this way we
should expect an abrupt transition from the weak disorder regime, at
small values of $\beta$, to the strong disorder regime, at large
values of $\beta$ \cite{Porto99}.

In order to quantify macroscopically the effect of disorder on the
geometry of the OPC, we performed computer simulations for $1000$
realizations of lattices with sizes varying in the range $32 \le L \le
512$, and generated with different values of the disorder parameter
$\beta$. In Fig.~2 we show for $\beta=0.002$ that the average mass of
the OPC backbone scales as $M_{b} \simeq L^{D_{b}}$, with an exponent
$D_{b}=1.22 \pm 0.02$ \cite{self-affine}. Surprisingly, this exponent
value is statistically identical to the fractal dimension previously
found for the optimal path line, but obtained under strong disorder
\cite{Cieplak94,Porto97,Porto99}. It is also very close to that found
``strands'' in Invasion Percolation ($1.22 \pm 0.01$~\cite{Cieplak94}), 
and paths on Minimum Spanning Trees ($1.22 \pm 0.01$~\cite{Dobrin01}). 
In our case, however, it is important to note that the value of
$D_{b}$ reflects a highly non-local property of the system that is
intrinsically associated with the iterative process involved in the
OPC calculation. Also shown in Fig.~2 is the mass of the OPC fracture
in weak disorder, which consists of both the backbone and its dangling
ends. While the crack itself grows as a power-law with size, $M_{f}
\simeq L^{D_{f}}$, with an exponent $D_{f}=1.59 \pm 0.02$, the total
mass of blocked sites (crack and isolated clusters), however, is a
constant fraction of the total mass of the system, i.e. $M_{t} \simeq
L^{2}$.
\begin{figure}[!htb]
\includegraphics*[width=8cm]{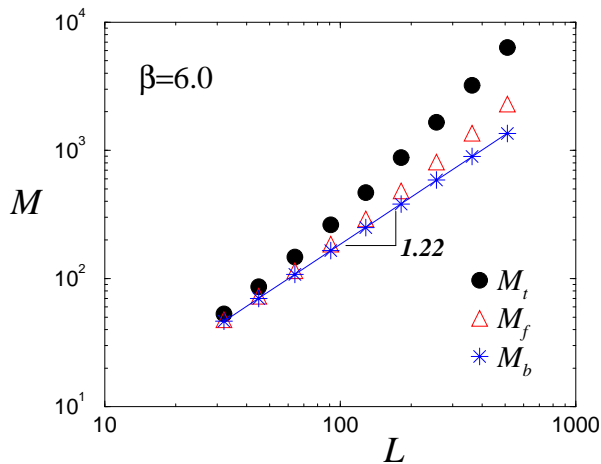}
\caption{(Color online) Log-log plot of the masses $M_{t}$ (circles), 
$M_{f}$ (triangles) and $M_{b}$ (stars) against system size $L$ for
$\beta=6.0$. For this intermediate value of $\beta$, we can clearly
identify the crossover from strong to weak disorder depending on the
system scale. For small lattices, the system operates under strong
disorder conditions. As a consequence, most of the blocked sites lie
on the fracture path and the three masses $M_{t}$, $M_{f}$ and $M_{b}$
are identical. As the system size increases, one reaches the weak
disorder regime, and the three curves split apart. At larger scales,
we should recover the same power-law behaviours found for sufficiently
low values of $\beta$, as shown in Fig.~2.}
\label{ws natural}
\end{figure}

In Fig.~3, the results obtained for $\beta=6.0$ clearly indicate the
transition from strong to weak disorder regimes by systematically
increasing the size of the system. As already mentioned, the stronger
the disorder in the system (low $L$ or high $\beta$), the smaller is
the number of final blocked sites that also become more and more
localized in a singly-connected crack line. Precisely, in the limit of
very strong disorder, we obtain that only the OPC backbone mass
$M_{b}$ remains, i.e. $M_{t} \rightarrow M_{b}$ and $M_{f} \rightarrow
M_{b}$, scaling in the same way as in the weak disorder limit, namely
$M_{b}\simeq L^{D_{b}}$, with $D_{b}=1.22 \pm 0.02$. As shown in
Fig.~1, the backbone is indeed invariant under the change of
$\beta$. By increasing the value of $L$, we observe a gradual
departure of $M_{t}$ and $M_{f}$ from this behavior to their
respective scaling laws in weak disorder (high $L$ or low $\beta$), as
displayed in Fig.~2.
\begin{figure}[!htb]
\includegraphics*[width=8cm]{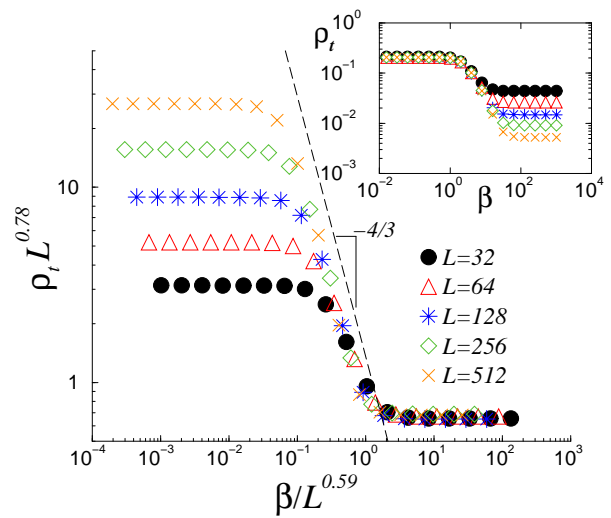}
\caption{(Color online) Transition from weak to strong disorder in the 
OPC model. In these curves we plot the density of all blocked sites
$\rho_{t}$ at the end of the process when the top of the system
disconnects from the bottom. These curves present three distinct
regimes. For weak disorder $\beta<1$ one sees a saturation in the
density in a value around $\rho_{t}\approx{0.22}$. For intermediate
values one sees the density decreasing with the disorder parameter
$\beta$, however there is no dependence on the system size. For larger
values of $\beta$ we see the density saturating again, but now to a
value that depends on the system size $L$. This last regime is the one
that characterizes the strong disorder regime. The saturated density
in this regime depends on the system size as $L^{D_{b}-2}$, where
$D_{b}\approx{1.22}$ is the fractal dimension of the OPC backbone. For
each system size, we observe the transition to the strong disorder
regime at a different value of the parameter $\beta_{\times}$. From
the collapse shown in the inset, a non-trivial dependence is revealed
between the onset of the transition and the system size,
$\beta_{\times}\sim{L^{0.59}}$.}
\end{figure}

The transition from weak to strong disorder is better illustrated by
the results depicted in the inset of Fig.~4. There we plot the density
of all blocked sites $\rho_{t}$ at the end of the process as a
function of the parameter $\beta$. The curves exhibit three different
regimes, depending on the value of $\beta$. For small values,
$\beta<1$, the density saturates to a fixed value. For larger values,
the density decays as a power-law $\rho_{t}\sim\beta^{-\theta}$, with
an exponent $\theta \approx 4/3$. Both regimes, saturation and
power-law decay, are still in weak disorder. For finite lattices, the
curves present another crossover to a minimum density that is now
dependent on the system size. This second crossover, $\beta_{\times}$,
which indicates the transition to strong disorder, should depend on
the system size in such way that an infinite system is in weak
disorder for any finite value of $\beta$. For large enough values of
$\beta$, in the strong disorder regime, the density reaches a minimum
value when all the blocked sites lie on the fracture dividing the
system. As shown in Fig.~2, the mass of this fracture scales as
$L^{D_{b}}$. At the onset of the transition, this power-law behavior
crosses over to a scale dependent value. Thus,
$\beta_{\times}^{-\theta}\sim{L^{D_{b}-2}}$, or
$\beta_{\times}\sim{L^{\alpha}}$, with $\alpha=(2-D_{b})/\theta$.  The
collapse of the results for intermediate and large values of $\beta$
obtained using $D_{b}=1.22$ and $\alpha=0.59$ shown in the main plot
of Fig.~4 is consistent with this analysis.

Concluding, we discover that for all disorders the line along which
all minimum energy paths fracture is fractal of dimension $1.22$ in
$2d$. The role of disorder and system size can be fully cast in a
crossover scaling law for the total number of blocked sites. Our model
poses new challenges also from the theoretical point of view since the
numerical resemblance of our fractal to the one of domain walls and
optimal paths in the strong disorder limit \cite{Cieplak94} seems to
hint towards some deeper relation. It could be certainly interesting
to study also the influence of the dimension of the system and of the
substrate topology on our model by a generalization to three dimensional
systems or complex network.
 
We thank the Brazilian agencies CAPES, CNPq, FUNCAP and FINEP, and 
the ETH Competence Center CCSS in Switzerland for financial support .

\bibliographystyle{prsty}

\begin{thebibliography}{25}
\bibitem{self-affine}
The roughness exponent found for the OPC bacbone is equal to unity
within the statistical error bars, supporting the fact that this is
indeed a self-similar object and not self-affine for intermediate to
large scales.
\end{thebibliography}

\end{document}